\documentclass[a4paper]{article}

\usepackage[english]{babel}
\usepackage[utf8x]{inputenc}
\usepackage[T1]{fontenc}

\usepackage[a4paper,top=3cm,bottom=2cm,left=3cm,right=3cm,marginparwidth=1.75cm]{geometry}

\usepackage{amsmath}
\usepackage{bbold}
\usepackage{authblk}
\usepackage{bm}
\usepackage{graphicx}
\usepackage{multirow}
\usepackage{caption}  % for figure caption
\usepackage{booktabs}
\usepackage{algorithm}
\usepackage{algpseudocode}
\providecommand{\keywords}[1]{\textbf{\textit{Keywords:}} #1}

\title{Logistic principal component analysis via non-convex singular value thresholding}

\author{Yipeng Song, Johan A. Westerhuis, Age K. Smilde}
\affil{Swammerdam Institute for Life Sciences, University of Amsterdam}
\date{}

\begin{document}

\maketitle

\begin{abstract}
Multivariate binary data is becoming abundant in current biological research. Logistic principal component analysis (PCA) is one of the commonly used tools to explore the relationships inside a multivariate binary data set by exploiting the underlying low rank structure. We re-expressed the logistic PCA model based on the latent variable interpretation of the generalized linear model on binary data. The multivariate binary data set is assumed to be the sign observation of an unobserved quantitative data set, on which a low rank structure is assumed to exist. However, the standard logistic PCA model (using exact low rank constraint) is prone to overfitting, which could lead to divergence of some estimated parameters towards infinity. We propose to fit a logistic PCA model through non-convex singular value thresholding to alleviate the overfitting issue. An efficient Majorization-Minimization algorithm is implemented to fit the model and a missing value based cross validation (CV) procedure is introduced for the model selection. Our experiments on realistic simulations of imbalanced binary data and low signal to noise ratio show that the CV error based model selection procedure is successful in selecting the proposed model. Furthermore, the selected model demonstrates superior performance in recovering the underlying low rank structure compared to models with convex nuclear norm penalty and exact low rank constraint. A binary copy number aberration data set is used to illustrate the proposed methodology in practice.
\end{abstract}


\section{Introduction}
Principal component analysis (PCA) is a canonical method to extract the low rank structure from a high dimensional multivariate quantitative data set \cite{jolliffe2002principal, bro2014principal}. The results derived from a PCA model can be used for exploratory data analysis or as input for other statistical methods. Current biological research has also seen an increasing abundance of high dimensional multivariate binary data sets. Examples include comprehensive point mutation, copy number aberration (CNA) and binarized methylation measurements \cite{cancer2008comprehensive, iorio2016landscape}. To tackle these multivariate binary data sets, the classical PCA model has been generalized from different perspectives to take into account the special mathematical properties of binary data \cite{de2009gifi, kiers1989three, collins2002generalization}. Results derived from these PCA extensions on multivariate binary data can be interpreted and used in a similar way as the classical PCA model.\\

Logistic PCA \cite{de2003principal, schein2003generalized} is one of the PCA extensions in the probabilistic framework. It is motivated from the probabilistic interpretation of the classical PCA model with Gaussian distributed error. The extension of the classical PCA model to the logistic PCA model is similar to the extension of linear regression to logistic linear regression. In the classical PCA model, the low rank constraint is imposed on the conditional mean of the observed quantitative data set, while in the logistic PCA model, the low rank constraint is imposed on the logit transform of the conditional mean of the observed binary data. Therefore, the logistic PCA model can also be re-expressed in a similar way as the latent variable interpretation of the generalized linear models (GLMs) on binary data \cite{agresti2013categorical}. In logistic PCA, the observed binary data set can be assumed as the sign observation of an unobserved quantitative data set, on which low rank structure is assumed to exist. This intuitive latent variable interpretation not only facilitates the understanding of the low rank structure in the logistic PCA model, but also provides a way to define the signal to noise ratio (SNR) in the simulation of multivariate binary data.\\

However, the standard logistic PCA model with the exact low rank constraint, which is expressed as the multiplication of two low rank matrices, is prone to overfitting, leading to divergence of some estimated parameters towards infinity \cite{de2003principal, groenen2016multinomial, song2017principal}. The same overfitting problem also happens for the logistic linear regression model. If two classes of the outcome are linearly separable with respect to an explanatory variable, the corresponding coefficient of this variable tends to go to infinity \cite{agresti2013categorical}. A common trick is adding a ridge regression (quadratic) type penalty on the coefficient vector to alleviate the overfitting issue. If we apply the same trick on the logistic PCA model, the quadratic penalty on the loading matrix is equivalent to a quadratic penalty on the singular values of a matrix, which is the multiplication of the score and loading matrices. Details will be shown later. Therefore, it is possible to derive a robust logistic PCA model via regularization of the singular values. \cite{davenport20141} proposed to use a nuclear norm penalty in the low rank matrix approximation framework for the binary matrix completion problem. The proposed method is similar to the logistic PCA model except that the column offset term is not included and the exact low rank constraint is replaced by its convex relaxation, the nuclear norm penalty. The nuclear norm penalty, which is equivalent to applying a lasso penalty on the singular values of a matrix, induces low rank estimation and constrains the scale of non-zeros singular values simultaneously. However, a lasso type penalty shrinks all parameters to the same degree, leading to biased parameter estimation. This behavior will further make the CV error or the prediction error based model selection procedure inconsistent \cite{meinshausen2010stability}. On the other hand, non-convex penalties, many of which are concave functions, are capable to simultaneously achieve nearly unbiased parameter estimation and sparsity \cite{fan2001variable,armagan2013generalized}. Recent research \cite{shabalin2013reconstruction, josse2016adaptive} has also shown the superiority of non-convex singular value thresholding (applying non-convex penalties on the singular values of a matrix) in recovering the true signal in a low rank approximation framework under Gaussian noise. In this paper, we propose to fit the logistic PCA model via non-convex singular value thresholding as a way to alleviate the overfitting problem and to induce low rank estimation simultaneously. A Majorization-Minimization (MM) algorithm is implemented to fit the proposed model and an option for missing values is included. In the developed algorithm, the updating of all the parameters has an analytical form solution, and the loss function is guaranteed to decrease in each iteration. After that, a missing value based cross validation procedure is introduced for the model selection.\\

Based on the latent variable interpretation of the logistic PCA model, realistic multivariate binary data sets (low SNR, imbalanced binary data) are simulated to evaluate the performance of the proposed model and the corresponding model selection procedure. It turns out that the CV error based model selection procedure is successful in the selection of the proposed model, and the selected model has superior performance in recovering the underlying low rank structure compared to the model with convex nuclear norm penalty and exact low rank constraint. Furthermore, the performance of the logistic PCA model as a function of the SNR in multivariate binary data simulation is fully characterized. Finally, a binary CNA data set is used to illustrate the proposed methodology in practise.\\

\section{Latent variable interpretation of models on binary data}
\subsection{Latent variable interpretation of the GLMs on binary data}
A univariate binary response variable $y$ is assumed to follow a Bernoulli distribution with parameter $\pi$, $y \sim \text{Bernoulli}(\pi)$. $\mathbf{x}$ is a multivariate explanatory variable and $\mathbf{x} \in \mathbf{R}^J$. For the GLMs on binary data, we assume that the nonlinear transformation of the conditional mean of $y$ is a linear function of $\mathbf{x}$, $h(\text{E}(y|\mathbf{x})) = \mathbf{x}^{\text{T}} \bm{\beta}$, in which $h()$ is the link function, $\text{E}(y|\mathbf{x})$ is the conditional mean, and $\bm{\beta}$ is a $J$ dimensional coefficient vector. If the inverse function of $h()$ is $\phi()$, we have $\text{E}(y|x) = \phi(x^{\text{T}}\bm{\beta})$. If the logit link is used, $\phi(\theta) = (1+\exp(-\theta))^{-1}$, which is the logistic linear regression model, and $x^{\text{T}}\bm{\beta}$ can be interpreted as the log-odds, which is the natural parameter of Bernoulli distribution expressed in exponential family distribution form. If the probit link is used, $\phi(\theta) = \Phi(\theta)$, in which $\Phi(\theta)$ is the cumulative density function (CDF) of the standard normal distribution, which is the probit linear regression model.\\

The fact that the inverse link function $\phi()$ can be interpreted as the CDF of a specific probability distribution, motivates the latent variable interpretation of the logistic or probit linear regression \cite{agresti2013categorical}. $y$ can be assumed as the sign observation of a quantitative latent variable $y^{\ast}$, which has a linear relationship with the explanatory variable $\mathbf{x}$. Taking the probit linear regression as an example, the latent variable interpretation can be expressed as,
\begin{equation*}
\begin{split}
        y^{\ast} &= \mathbf{x}^{\text{T}}\beta + \epsilon\\
 \epsilon &\sim \text{N}(0,1) \\
        y &= \mathbb{1}{(y^{\ast}>0)},
\end{split}
\end{equation*}
in which $y^{\ast}$ is the latent variable, $\epsilon$ is the error term, and $\mathbb{1}()$ is the indicator function. The probability for the observation $y = 1$ is $\text{Pr}(y=1|\mathbf{x}^{\text{T}}\bm{\beta}) = \text{Pr}(y^{\ast} \geq 0)= \Phi(\mathbf{x}^{\text{T}}\bm{\beta})$. A similar latent variable interpretation can be applied to the logistic linear regression model by assuming that the error term $\epsilon$ follows the standard logistic distribution. The probability density function of the logistic distribution can be expressed as,
\begin{equation*}
  p(\epsilon) = \frac{\exp(- \frac{\epsilon-\mu}{\sigma})}{\sigma(1+\exp(-\frac{\epsilon-\mu}{\sigma}))^2},
\end{equation*}
in which $\mu$ and $\sigma$ are the location and scale parameters. In the standard logistic distribution, $\mu=0$, $\sigma=1$. The inverse-logit function $\phi()$ is the CDF of the standard logistic distribution. The assumption of $\mu=0$ for the $\epsilon$ is reasonable since we want to use the linear function $\mathbf{x}^{\text{T}}\bm{\beta}$ to capture the conditional mean of $y^{\ast}$. The assumption of $\sigma=1$ for the $\epsilon$ seems restrictive, however scaling the estimated $\hat{\bm{\beta}}$ by a positive constant as $\hat{\bm{\beta}}/\sigma$ will not change the conclusion of the model. Since the assumption of logistic distributed noise is not very straightforward, the latent variable interpretation of the logistic linear regression model is not widely used.\\

The above latent variable interpretation of the GLMs on binary data is naturally connected to the generating process of binary data \cite{young1980quantifying}. Binary data can be discrete in nature, for example when females and males are classified as ``1'' and ``0''. Another possibility is that there is a continuous process underlying the binary observation. For example in a toxicology study, the binary outcome of a subject being dead or alive relates to the dosage of a toxin used and the subject's tolerance level. The tolerance varies for different subjects, and the status (dead or alive) of a specific subject depends on whether its tolerance is higher than the used dosage or not. Thus, a continuous tolerance level is underlying the binary outcome \cite{agresti2013categorical}. If we assume our binary data set is generated from a continuous process, it is natural to use the latent variable interpretation of the probit link; or if it is assumed from a discrete process, we can use the logit link, and interpret it from the probabilistic perspective rather than the latent variable perspective. However, usually, the difference between the results derived out from the GLMs using logit or probit link is negligible \cite{agresti2013categorical}.\\

\subsection{Latent variable interpretation of the logistic PCA model}
The measurement of $J$ binary variables on $I$ samples results in a binary matrix $\mathbf{X}$($I\times J$), whose $ij$-th element $x_{ij}$ equals ``1'' or ``0''. The logistic PCA model on $\mathbf{X}$ can be interpreted as follows. Conditional on the low rank structure assumption, which is used to capture the correlations observed in $\mathbf{X}$, elements in $\mathbf{X}$ are independent realizations of the Bernoulli distributions, whose parameters are the corresponding elements of a probability matrix $\mathbf{\Pi}$($I \times J$), $\text{E}(\mathbf{X}|\mathbf{\Pi}) = \mathbf{\Pi}$. Assuming the natural parameter matrix, which is the logit transform of the probability matrix $\mathbf{\Pi}$, is $\mathbf{\Theta}$($I \times J$), we have $h(\mathbf{\Pi}) = \mathbf{\Theta}$ and $\mathbf{\Pi} = \phi(\mathbf{\Theta})$, in which $h()$ and $\phi()$ are the element-wise logit and inverse logit functions. The low rank structure is imposed on $\mathbf{\Theta}$ in the same way as in a classical PCA model, $\mathbf{\Theta} = \mathbf{1}\bm{\mu}^{\text{T}} + \mathbf{A}\mathbf{B}^{\text{T}}$, in which $\bm{\mu}$($J\times 1$) is the $J$ dimensional column offset term and can be interpreted as the logit transform of the marginal probabilities of the binary variables. $\mathbf{A}$ ($I \times R$) and $\mathbf{B}$($J \times R$) are the corresponding low rank score and loading matrices, and $R$, $R \ll \text{min}(I,J)$, is the low rank. Therefore, for the logistic PCA model, we have $\text{E}(\mathbf{X}|\mathbf{\Theta}) = \phi(\mathbf{\Theta}) = \phi(\mathbf{1}\bm{\mu}^{\text{T}} + \mathbf{A}\mathbf{B}^{\text{T}})$. On the other hand, in a classical PCA model, we have $\text{E}(\mathbf{X}|\mathbf{\Theta}) = \mathbf{\Theta} = \mathbf{1}\bm{\mu}^{\text{T}} + \mathbf{A}\mathbf{B}^{\text{T}}$, which is equivalent to using the identity link function. Furthermore, unlike in the classical PCA model, the column offset $\bm{\mu}$ has to be included into the logistic PCA model to do the model based column centering. The reason is that the commonly used column centering processing step is not allowed to be applied on the binary data set as the column centered binary data is not binary anymore.\\

The logistic PCA model can be re-expressed in the same way as the latent variable interpretation of the GLMs on binary data. Our binary observation $\mathbf{X}$ is assumed to be the sign observation of an underlying quantitative data set $\mathbf{X}^{\ast}$($I\times J$), and for the $ij$-th element, we have $x_{ij} = 1$ if  $x^{\ast}_{ij} \geq 0$ and $x_{ij} = 0$ \textit{vice versa}. The low rank structure is imposed on the latent data set $\mathbf{X}^{\ast}$ as $\mathbf{X}^{\ast} = \mathbf{\Theta} + \mathbf{E}$, in which $\mathbf{E}$($I\times J$) is the error term, and its elements follow a standard logistic distribution. The latent variable interpretation of the logistic PCA model can be expressed as,
\begin{equation*}
\begin{split}
                  \mathbf{X}^{\ast} &= \mathbf{\Theta} + \mathbf{E} \\
                  \epsilon_{ij} & \sim \text{Logistic}(0,1), i = 1 \cdots I, j = 1 \cdots J \\
                  x_{ij} & = \mathbb{1}{(x^{\ast}_{ij}>0)}, i = 1 \cdots I, j = 1 \cdots J.
\end{split}
\end{equation*}

Similar to the latent variable interpretation of the logistic linear model, the assumption of $\epsilon_{ij} \sim \text{Logistic}(0,1)$ is not restrictive, since scaling the estimated $\hat{\mathbf{\Theta}}$ by a positive constant $\sigma$ will not change the conclusions from the model. When the standard normal distributed error is used in the above derivation, we get the probit PCA model. The latent variable interpretation of the logistic or probit PCA not only facilitates our understanding of the low rank structure underlying a multivariate binary data, but also provides a way to define the SNR in multivariate binary data simulation.\\

\section{Logistic PCA model via singular value thresholding}
\subsection{The standard logistic PCA model}
Assume the column centered $\mathbf{\Theta}$ is $\mathbf{Z}$, $\mathbf{Z} = \mathbf{\Theta} - \mathbf{1}\bm{\mu}^{\text{T}} = \mathbf{AB}^{\text{T}}$. In the standard logistic PCA model, the exact low rank constraint is imposed on $\mathbf{Z}$ as the multiplication of two rank $R$ matrices $\mathbf{A}$ and $\mathbf{B}$. The negative log likelihood of fitting the observed $\mathbf{X}$ conditional on the low rank structure assumption on $\mathbf{\Theta}$ is used as the loss function. We also introduce a weighting matrix $\mathbf{W}$($I \times J$) to tackle the potential missing values in $\mathbf{X}$. The $ij$-th element of $\mathbf{W}$, $w_{ij}$, equals 0 when the corresponding element in $\mathbf{X}$ is missing; while it is 1 \textit{vice versa}. The optimization problem of the standard logistic PCA model can be expressed as,
\begin{equation}\label{eq1}
\begin{aligned}
\min_{\bm{\mu}, \mathbf{Z}} \quad & -\log(p(\mathbf{X}|\mathbf{\Theta},\mathbf{W}))\\
               &= -\log(\prod_{i}^{I}\prod_{j}^{J} (p(x_{ij}|\theta_{ij}))^{w_{ij}})\\
               &= -\sum_{i}^{I}\sum_{j}^{J} w_{ij} \left[x_{ij}\log(\phi(\theta_{ij})) + (1-x_{ij})\log(1-\phi(\theta_{ij}))\right] \\
           \text{s.t.} \quad   &\mathbf{\Theta} = \mathbf{1}\bm{\mu}^{\text{T}} + \mathbf{Z}\\
                               &\text{rank}(\mathbf{Z}) = R \\
                               &\mathbf{1}^{\text{T}}\mathbf{Z} = \mathbf{0},
\end{aligned}
\end{equation}
in which the constraint $\mathbf{1}^{\text{T}}\mathbf{Z} = \mathbf{0}$ is imposed to make $\bm{\mu}$ identifiable. Unfortunately, the classical logistic PCA model tends to overfit the observed binary data. In order to decrease the loss function in equation \ref{eq1}, $\theta_{ij}$ tends to approach positive infinity when $x_{ij}=1$, and negative infinity when $x_{ij} = 0$. This overfitting problem will be explored in more detail below. In logistic linear regression, this overfitting problem can be solved by adding a quadratic penalty on the coefficient vector to regularize the estimated parameters. A similar idea can be applied to the logistic PCA model by taking it as a regression type problem. The columns of the score matrix $\mathbf{A}$ are taken as the unobserved explanatory variables, while the loading matrix $\mathbf{B}$ are the coefficients. If we decompose $\mathbf{Z}$ into a $R$ truncated SVD as $\mathbf{Z}=\mathbf{UDV}^{\text{T}}$, then $\mathbf{A}=\mathbf{U}$ and $\mathbf{B}=\mathbf{VD}^{\text{T}}$. It is easy to show that the quadratic penalty $||\mathbf{B}||_F^2 = \sum_{r} \sigma_{r}^2$, in which $\sigma_{r}$ is the $r$-th singular value of $\mathbf{Z}$. Therefore, it is possible to derive a robust logistic PCA model by thresholding the singular values of $\mathbf{Z}$.\\

\subsection{Logistic PCA via non-convex singular value thresholding}
The most commonly used penalty function in thresholding singular values is the nuclear norm penalty, and it has been used in solving many low rank approximation problems \cite{candes2009exact,mazumder2010spectral,davenport20141,groenen2016multinomial}. If the SVD decomposition of matrix $\mathbf{Z}$ is $\mathbf{Z}=\mathbf{UDV}^{\text{T}}$, the nuclear norm penalty can be expressed as $\sum_{r} \sigma_r$, in which $\sigma_r$ is the $r$-th singular value. The nuclear norm penalty is the convex relaxation of the exact low rank constraint and can be regarded as applying a lasso penalty on the singular values of a matrix. Therefore, the nuclear norm penalty has the same problem as the lasso penalty, it shrinks all singular values to the same degree. This leads to a biased estimation of the large singular values. This behavior will further make the prediction error or cross validation error based model selection procedure inconsistent \cite{meinshausen2010stability}. As an alternative, non-convex penalties can shrink the parameters in a nonlinear manner to achieve nearly unbiased estimation and provide sparse solutions simultaneously \cite{fan2001variable,armagan2013generalized}. Therefore, we propose to replace the exact low rank constraint in the logistic PCA model by a concave penalty on the singular values of $\mathbf{Z}$ to achieve a low rank estimation continuously and to alleviate the overfitting issue. In this paper, we use the frequentist version of the generalized double Pareto (GDP) \cite{armagan2013generalized} penalty as an example of the concave penalties to show the results. We also provide the option of the smoothly clipped absolute deviation (SCAD) penalty \cite{fan2001variable} and $L_{q:0<q \leq 1}$ penalty \cite{fu1998penalized} in our implementation. The GDP penalty on the singular values of $\mathbf{Z}$ is $g(\mathbf{Z}) = \sum_{r} g(\sigma_r) = \sum_{r} \log(1+\frac{\sigma_{r}}{\gamma})$, in which $g()$ is the concave function used, $\sigma_{r}$ is the $r$-th singular value of $\mathbf{Z}$ and $\gamma$ is the hyper-parameter of the penalty function. The thresholding properties of the exact low rank constraint, the nuclear norm penalty and the GDP penalty with different values of $\gamma$ are shown in Fig.~1. The penalized negative log likelihood for fitting the observed binary data $\mathbf{X}$ of the logistic PCA with GDP penalty can be shown as,
\begin{equation}\label{eq2}
\begin{aligned}
\min_{\bm{\mu}, \mathbf{Z}} \quad & -\log(p(\mathbf{X}|\mathbf{\Theta},\mathbf{W})) + \lambda g(\mathbf{Z}) \\
           \text{s.t.} \quad   &\mathbf{\Theta} = \mathbf{1}\bm{\mu}^{\text{T}} + \mathbf{Z}\\
                               &\mathbf{1}^{\text{T}}\mathbf{Z} = \mathbf{0},
\end{aligned}
\end{equation}
in which $\log(p(\mathbf{X}|\mathbf{\Theta},\mathbf{W}))$ and $g(\mathbf{Z})$ are as described above.\\

\begin{figure}[h!]\label{Fig:1}
    \centering
    \includegraphics[width=0.5\textwidth ]{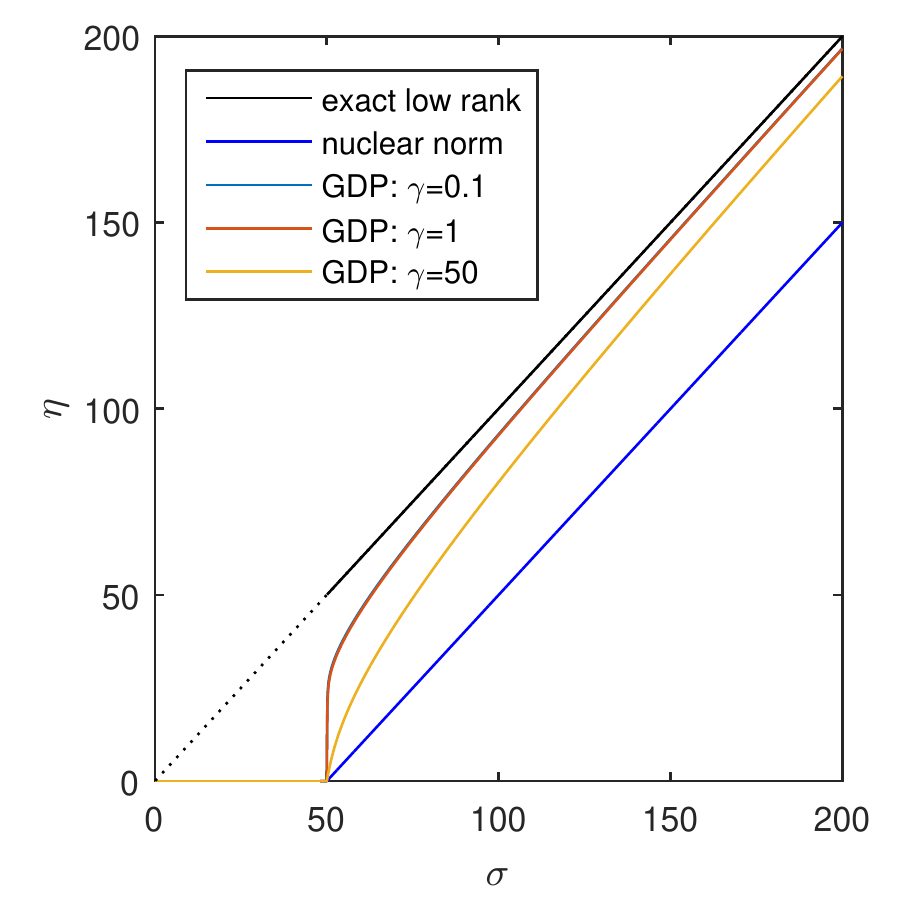}
    \caption*{\textbf{Fig.~1} Thresholding properties of the exact low rank constraint, the nuclear norm and the GDP penalty with different values of $\gamma$. Note that the curves corresponding to the GDP penalty with $\gamma=1$ and $\gamma=0.1$ almost perfectly overlap. $\sigma$: nonnegative singular values; $\eta$: singular values after thresholding.}
\end{figure}

\section{Algorithm}
Based on the Majorization-Minimization (MM) principle \cite{de1994block,hunter2004tutorial}, an MM algorithm is derived to fit the logistic PCA model via non-convex singular value thresholding. The derived algorithm is guaranteed to decrease the objective function in equation 2 during each iteration and the analytical form for updates of all the parameters in each iteration exist. Although the following derivation focuses on using the logit link function and the GDP penalty, the option for probit link, $L_{q}$ penalty and SCAD penalty are included in our implementation.\\

\subsection{The majorization of the penalized negative log-likelihood}
The negative log-likelihood is $f(\mathbf{\Theta}) = -\log(p(\mathbf{X}|\mathbf{\Theta},\mathbf{W}))$, and the concave penalty is $g(\mathbf{Z}) = \sum_{r} g(\sigma_r)$. $f(\mathbf{\Theta})$ can be majorized to a quadratic function of $\mathbf{\Theta}$ by exploiting the upper-bound of the second order gradient of $f(\mathbf{\Theta})$, while the penalty $g(\mathbf{Z})$ can be majorized to a linear function of the singular values of $\mathbf{Z}$ by exploiting the property of a concave function.  The derivation process is the same as the algorithm in our previous research \cite{song2018generalized}, therefore we will only show the result here.

\begin{equation}\label{eq3}
\begin{aligned}
f(\mathbf{\Theta}) &\leq \frac{L}{2}||\mathbf{\Theta}-\mathbf{H}^k||_F^2 + c\\
g(\mathbf{Z}) &\leq \sum_{r} \nabla g(\sigma_r^k)\sigma_{r} + c\\
              \mathbf{H}^k &= \mathbf{\Theta}^k - \frac{1}{L} (\mathbf{W}\odot \nabla f(\mathbf{\Theta}^k))\\
\nabla f(\mathbf{\Theta}^k)) &= \phi(\mathbf{\Theta}^{k}) - \mathbf{X}\\
\nabla g(\sigma_r^k) &=\frac{1}{\gamma+\sigma_r^k},
\end{aligned}
\end{equation}
in which $L$ is the upper-bound of the second order gradient of $f(\mathbf{\Theta})$, and can always set to $L=0.25$; $\sigma_r^k$ is the $r$-th singular value of $\mathbf{Z}^{k}$, which is an approximation of $\mathbf{Z}$ during the $k$-th iteration; $\mathbf{\Theta}^{k}$ is the approximation of $\mathbf{\Theta}$ during the $k$-th iteration; $\phi()$ is the inverse logit function; $\nabla f(\mathbf{\Theta}))$ and $\nabla g(\sigma_r)$ are the first order gradients of $f(\mathbf{\Theta}))$ and $g(\sigma_r)$ respectively. Summarizing these two majorization steps, we have the following majorized problem during the $k$-th iteration.
\begin{equation}\label{eq4}
\begin{aligned}
\min_{\bm{\mu},\mathbf{Z}} \quad & \frac{L}{2}||\mathbf{\Theta}-\mathbf{H}^{k}||_F^2 + \lambda \sum_{r} \nabla g(\sigma_r^k)\sigma_{r}\\
           \text{s.t.} \quad   &\mathbf{\Theta} = \mathbf{1}\bm{\mu}^{\text{T}} + \mathbf{Z}\\
                               &\mathbf{1}^{\text{T}}\mathbf{Z} = \mathbf{0} \\
                               & \mathbf{H}^k = \mathbf{\Theta}^k - \frac{1}{L} (\mathbf{W}\odot \nabla f(\mathbf{\Theta}^k)).
\end{aligned}
\end{equation}

\subsection{Block coordinate descent}
The majorized problem in equation \ref{eq4} during the $k$-th iteration can be solved by the block coordinate descent algorithm.\\

\subsubsection*{Updating $\bm{\mu}$}
When fixing $\mathbf{Z}$ in equation \ref{eq4}, the analytical form solution of $\bm{\mu}$ is the column mean of $\mathbf{H}^k$, $\bm{\mu} = \frac{1}{I} (\mathbf{H}^k)^{\text{T}} \mathbf{1}$.\\

\subsubsection*{Updating $\mathbf{Z}$}
After deflating the offset set term $\bm{\mu}$ in equation \ref{eq4}, the optimization problem of $\mathbf{Z}$ becomes $\min_{\mathbf{Z}} \frac{L}{2}||\mathbf{Z}-\mathbf{J}\mathbf{H}^k||_F^2 + \lambda \sum_{r} \nabla g(\sigma_r^k)\sigma_{r}$, in which $\mathbf{J}$ is the column centering operator $\mathbf{J} = \mathbf{I} - \frac{1}{I}\mathbf{11}^{\text{T}}$. This optimization problem is equivalent to finding the proximal operator of the weighted sum of singular values, for which the analytical form global solution exists \cite{lu2015generalized}. If the SVD decomposition of $\mathbf{J}\mathbf{H}^k$ is $\mathbf{J}\mathbf{H}^k = \mathbf{USV}^{\text{T}}$, the analytical form solution of $\mathbf{Z}$ is $\mathbf{Z} = \mathbf{U}\mathbf{S_{z}} \mathbf{V}^{\text{T}}$, in which $\mathbf{S_{z}} = \text{Diag}\{\text{max}(0, s_{r}-\frac{\lambda \nabla g(\sigma_r^k)}{L}) \}$, and $s_{r}$ is $r$-th element of the diagonal of $\mathbf{S}$.\\

\subsubsection*{Initialization}
$\mathbf{Z}^0$ and $\mu^0$ can be set according to the user imputed values, or by using the following random initialization strategy. All the elements in $\mathbf{Z}^0$ can be sampled from the standard uniform distribution and $\mu^0$ can be set to $\mathbf{0}$. In the following algorithm, $f^k$ indicates the objective value in equation \ref{eq2} during the $k$-th iteration, the relative change of the objective value is used as the stopping criteria. $\epsilon_f$ indicates the tolerance for the relative change of the objective value. Pseudocode of the algorithm described above is shown in Algorithm 1. \\

\begin{algorithm}[htb]
  \caption{An MM algorithm to fit the logistic PCA model via non-convex singular value thresholding.}
  \label{alg1}
  \begin{algorithmic}[1]
    \Require
      $\mathbf{X}$, $\lambda$, $\gamma$;
    \Ensure
      $\bm{\mu}$, $\mathbf{A}$, $\mathbf{B}$;
    \State $k = 0$;
    \State Compute $\mathbf{W}$ for missing values;
    \State Initialize $\mu^0$, $\mathbf{Z}^0$;

    \While{$(f^{k-1}-f^{k})/f^{k-1}>\epsilon_f$}
        \State $\mathbf{\Theta}^k = \mathbf{1}(\bm{\mu}^k)^{\text{T}} + \mathbf{Z}^k$;
        \State $\nabla f(\mathbf{\Theta}^{k})= \phi(\mathbf{\Theta}^k)-\mathbf{X}$;
        \State $\mathbf{H}^k = \mathbf{\Theta}^k - \frac{1}{L} (\mathbf{W}\odot \nabla f(\mathbf{\Theta}^k))$;
        \State $\nabla g(\sigma_r^k) = \frac{1}{\gamma+\sigma_r^k} $;
        \State $\bm{\mu}^{k+1} = \frac{1}{I}(\mathbf{H}^{k})^{\text{T}} \mathbf{1}$;
        \State $\mathbf{JH}=\mathbf{J}\mathbf{H}^{k}$;
        \State $\mathbf{USV}^{\text{T}} = \mathbf{JH}$;
        \State $S_{z} = \text{Diag}\{\text{max}(0, s_{r}-\frac{\lambda \nabla g(\sigma_r^k)}{L}) \}$;
        \State $\mathbf{Z}^{k+1} = \mathbf{U}\mathbf{S_{z}}\mathbf{V}^{\text{T}}$;
        \State $\mathbf{\Theta}^{k+1} = \bm{\mu}^{k+1} + \mathbf{Z}^{k+1}$;
        \State $k=k+1$;
    \EndWhile
    \State $\mathbf{A} = \mathbf{U}$;
    \State $\mathbf{B} = \mathbf{V}\mathbf{S_{z}}$;
  \end{algorithmic}
\end{algorithm}

\section{Real data set and simulation process}
\subsection{Real data set}
 The 410 multivariate binary copy number copy aberration (CNA) measurements on 160 cell lines with three cancer types (breast, lung and skin cancer types) from the Genomic Determinants of Sensitivity in Cancer 1000 (GDSC1000) \cite{iorio2016landscape}, are used in this paper as an example of biological binary data. In the binarized CNA measurement, ``1'' indicates aberration (gains or losses of a segment in chromosomal regions) occurred and ``0'' indicates the normal wild types status. The CNA data set is quite sparse, as on average there are about $6.7\%$ ``1'' in the whole data set. The sparse pattern and the empirical marginal probabilities of the CNA data set are visualized in the Fig.~S1.\\

\subsection{Simulation process}
A multivariate binary data set $\mathbf{X}$ is simulated according to the latent variable interpretation of the logistic PCA model. The SNR is defined as $\text{SNR}=\frac{||\mathbf{Z}||_F^2}{||\mathbf{E}||_F^2}$, in which $\mathbf{E}$ is the error term, and its elements are sampled from the standard logistic distribution. The column offset term $\bm{\mu}$ represents the logit transform of the marginal probabilities of the binary variables, and can be set to $\mathbf{0}$ to simulate balanced binary data or it can be set according to the characteristics of the real biological data set to simulate imbalanced binary data. The SVD of the rank $R$ matrix $\mathbf{Z}$ equals $\mathbf{Z}= \mathbf{UDV}^{\text{T}}$, in which $\mathbf{U}^{\text{T}}\mathbf{U} = \mathbf{I}_{R}$, $\mathbf{V}^{\text{T}}\mathbf{V} = \mathbf{I}_{R}$ and the diagonal of $\mathbf{D}$ contains the singular values. Elements in $\mathbf{U}$ and $\mathbf{V}$ are first sampled from $N(0,1)$. After that, the column mean of $\mathbf{U}$ is deflated to have $\mathbf{1}^{\text{T}}\mathbf{U}=\mathbf{0}$, and the SVD is used to force $\mathbf{U}$ being orthogonal. Also, $\mathbf{V}$ is forced to orthogonality by the Gram-Schmidt algorithm. Then, the diagonal matrix $\mathbf{D}_{pre}$, whose $R$ diagonal elements are the sorted absolute values of the samples from $N(1,0.5)$, is simulated. We express $\mathbf{D}$ as $\mathbf{D} = c\mathbf{D}_{pre}$, in which $c$ is a constant and is used to adjust the SNR in the simulation of the multivariate binary data. Now we have $\mathbf{\Theta} = \mathbf{1}\bm{\mu}^{\text{T}} + \mathbf{Z}$ according to the logistic PCA model and $\mathbf{X}^{\ast} = \mathbf{\Theta} + \mathbf{E}$ according to the latent variable interpretation. $\mathbf{\Theta}$ is transformed to a probability matrix $\mathbf{\Pi}$ by the inverse logit link function, $\mathbf{\Pi} = \phi(\mathbf{\Theta})$. Then, elements in $\mathbf{X}$ are independently sampled from a Bernoulli distribution, whose parameters are the corresponding elements of the probability matrix $\mathbf{\Pi}$. In this way, multivariate binary observation with low dimensional structure is generated.\\

\section{Model assessment and model selection}
\subsection{Model assessment}
The construction of a logistic PCA model of the simulated binary data $\mathbf{X}$ provides the estimated parameters $\hat{\bm{\mu}}$, $\hat{\mathbf{A}}$ and $\hat{\mathbf{B}}$, and $\hat{\mathbf{\Theta}} = \mathbf{1}\hat{\bm{\mu}}^{\text{T}} + \hat{\mathbf{AB}^{\text{T}}}$ and $\hat{\mathbf{\Pi}} = \phi(\hat{\mathbf{\Theta}})$ can be computed.
The model's ability in recovering the true $\mathbf{\Theta}$ can be evaluated by the relative mean squares error (RMSE), which is defined as $\text{RMSE}(\mathbf{\Theta}) = \frac{||\mathbf{\Theta}-\hat{\mathbf{\Theta}}||_F^2}{||\mathbf{\Theta}||_F^2}$, where $\mathbf{\Theta}$ is the true parameter.
The RMSEs in estimating $\bm{\mu}$ and $\mathbf{Z}$ are defined in the same way. In some cases the mean Hellinger distance (MHD) to quantify the similarity between the true probability matrix $\mathbf{\Pi}$ and the estimated $\hat{\mathbf{\Pi}}$ is used. Hellinger distance \cite{le2012asymptotics} is a symmetric measure to quantify the similarity between two probability distributions. Assuming the parameter of a Bernoulli distribution is $\pi$ and its estimation is $\hat{\pi}$, the Hellinger distance is defined as $\text{HD}(\pi,\hat{\pi}) = \frac{1}{\sqrt{2}}\sqrt{(\sqrt{\pi}-\sqrt{\hat{\pi}})^2 + (\sqrt{1-\pi}-\sqrt{1-\hat{\pi}})^2}$. The mean Hellinger distance between the probability matrix $\mathbf{\Pi}$ and its estimate $\hat{\mathbf{\Pi}}$ is defined as $\text{MHD}(\mathbf{\Pi}) =  \frac{1}{I\times J} \sum_{i,j}^{I,J}\text{HD}(\pi_{ij},\hat{\pi}_{ij})$.\\

\subsection{Model selection}
For the model selection on real data, a missing value based cross validation procedure is proposed. The CV error is computed as follows. First, elements in $\mathbf{X}$ are split into the training and test sets as follows: $10\%$ ``1''s and ``0''s of $\mathbf{X}$ are randomly selected as the test set $\mathbf{X}^{\text{test}}$, which are set to missing values, and the resulting data set is taken as $\mathbf{X}^{\text{train}}$. After getting an estimation of $\hat{\mathbf{\Theta}}$ from a logistic PCA on the $\mathbf{X}^{\text{train}}$, we can index out the elements, which are corresponding to the test set $\mathbf{X}^{\text{test}}$, as $\hat{\mathbf{\Theta}}^{\text{test}}$. Then the CV error is defined as the negative log-likelihood of using $\hat{\mathbf{\Theta}}^{\text{test}}$ to fit $\mathbf{X}^{\text{test}}$.\\

There are two tuning parameters, $\gamma$ and $\lambda$, during the model selection of the logistic PCA model with GDP penalty. However, the performance of the model is rather insensitive to the selection of $\gamma$, which will be shown below. After fixing the value of $\gamma$, we can use a grid search to select a proper value of $\lambda$ based on the minimum CV error. First, a sequence of $\lambda$ values can be selected from a proper searching range, after which logistic PCA models will be fitted with the selected $\lambda$ values on the training set $\mathbf{X}^{\text{train}}$. A warm start strategy, using the results of a previous model as the initialization of the next model, is used to accelerate the model selection process. The model with the minimum CV error is selected and then it is re-fitted on the full data set $\mathbf{X}$. Because the proposed model is non-convex and its result is sensitive to the used initialization, it is better to use the results derived from the selected model as the initialization of the model to fit the full data sets.\\

\section{Results}
\subsection{The standard logistic PCA model tends to overfit the data}
In this first section we will use the CNA data as an example. The algorithm from \cite{de2003principal} is implemented to fit the standard logistic PCA model. Constraints, $\mathbf{A}^{\text{T}}\mathbf{A} = \mathbf{I}$ and $\mathbf{1}^{\text{T}}\mathbf{A}$, are imposed. Two different standard logistic PCA models  are constructed of the CNA data, both with three components. The first model is obtained with low precision (stopping criteria was set to $\epsilon_f=10^{-4}$) while for the other model a high precision was used ($\epsilon_f=10^{-8}$). The initialization was the same for these two models. The low precision model converged already after 220 iterations, while the high precision model did not convergence even after 50000 iterations. The difference between the final objective values of these two models is not large, $8.12e+03$ and $7.58e+03$ respectively. However, as shown in Fig.~2, the scale of the loading plots derived from these two models is very different. When a high precision stopping criteria is used, some of the elements from the estimated loading matrix from the standard logistic PCA model tend to become very large.\\

\begin{figure}[h!]\label{Fig:2}
    \centering
    \includegraphics[width=\textwidth ]{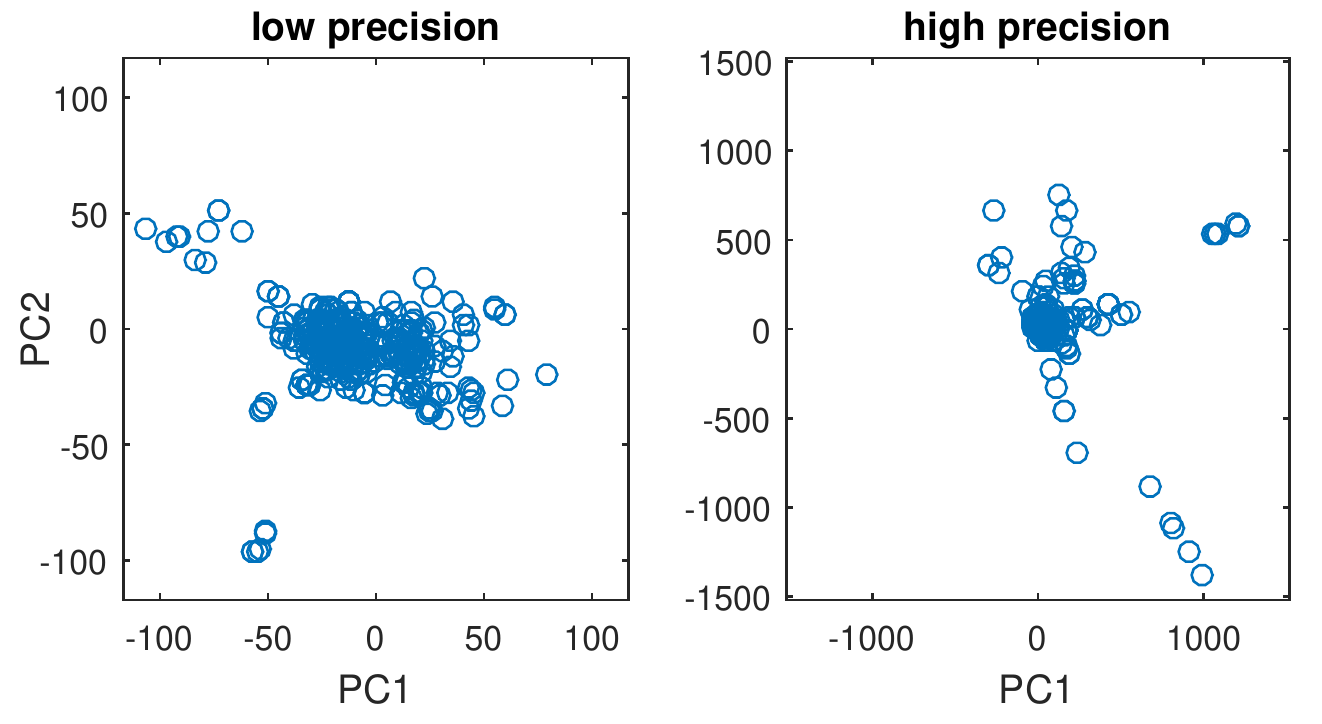}
    \caption*{\textbf{Fig.~2} The loading plots of the first two components derived from the low precision (left) and high precision (right) standard logistic PCA models.}
\end{figure}

\subsection{Model selection of the logistic PCA model with a GDP penalty}
A simulated data set is used here to show the CV error based model selection procedure. The offset term $\bm{\mu}$ is set to the logit transform of the empirical marginal probabilities of the CNA data to simulate an imbalanced binary data set. Other parameters used in the simulation are $I=160$, $J=410$, $\text{SNR}=1$ and $R=5$. First we will show the model selection procedure of $\lambda$ while the hyper-parameter $\gamma$ is fixed to $\gamma=1$. After splitting the simulated binary data set $\mathbf{X}$ into the training set $\mathbf{X}^{\text{train}}$ and the test set $\mathbf{X}^{\text{test}}$, 30 $\lambda$ values are selected from the searching range $[10,5000]$ with equal distance in log-space. For each $\lambda$ value, a logistic PCA with a GDP penalty ($\epsilon_f = 10^{-6}$, maximum iteration is 500) is constructed on $\mathbf{X}^{\text{train}}$ and for each model we evaluate its performance in estimating the simulated parameters. As shown in the model selection results (Fig.~3), the selected model with minimum CV error can also achieve approximately optimal RMSEs in estimating the simulated $\mathbf{\Theta}$, $\mathbf{Z}$ and $\bm{\mu}$ . However, the rank of the estimated $\mathbf{Z}$ from the selected model is 3, which is different from the simulated rank $R=5$. The reason will be discussed later. The selected model is re-fitted on the full simulated data $\mathbf{X}$, and the RMSEs of estimating $\mathbf{\Theta}$, $\mathbf{Z}$ and $\bm{\mu}$ are 0.0797, 0.2064 and 0.0421 respectively.\\

Next, we will show the model selection process of both $\gamma$ and $\lambda$. The simulated data $\mathbf{X}$ is split into the $\mathbf{X}^{\text{train}}$ and the $\mathbf{X}^{\text{test}}$ in the same way as described above. 30 $\gamma$ values are selected from the range $[10^{-1},10^{2}]$ equidistant in log-space. For each $\gamma$, 30 values of $\lambda$ are selected from a proper searching range, which is determined by an automatic procedure. For each value of $\gamma$, the model selection of $\lambda$ is done on the $\mathbf{X}^{\text{train}}$ in the same way as described above, after which the selected model is re-fitted on the full data $\mathbf{X}$. Therefore, for each value of $\gamma$, we have a selected model, which is optimal with respect to the CV error. As shown in Fig.~4(left), the difference between the RMSEs derived from these selected models is very small. This can be caused by two reasons: the model is insensitive to the selection of $\gamma$ or the CV error based model selection procedure is not successful in the selecting $\gamma$. To clarify the correct reason, we also fit $30 \times 30$ models on the full data $\mathbf{X}$ in the same way as the above experiment. For each value of $\gamma$, the model with minimum $\text{RMSE}(\mathbf{\Theta})$ is selected. As shown in Fig.~4(right), the value of $\gamma$ does not have a large effect on the RMSEs of the selected models, which are optimal with respect to the $\text{RMSE}(\mathbf{\Theta})$. Therefore, it can be concluded that the performance of the model is insensitive to the model selection of $\gamma$. Therefore, the strategy can be to  set a default value for $\gamma$ and focus on the selection of $\lambda$.\\

\begin{figure}[h!]\label{Fig:3}
    \centering
    \includegraphics[width=\textwidth ]{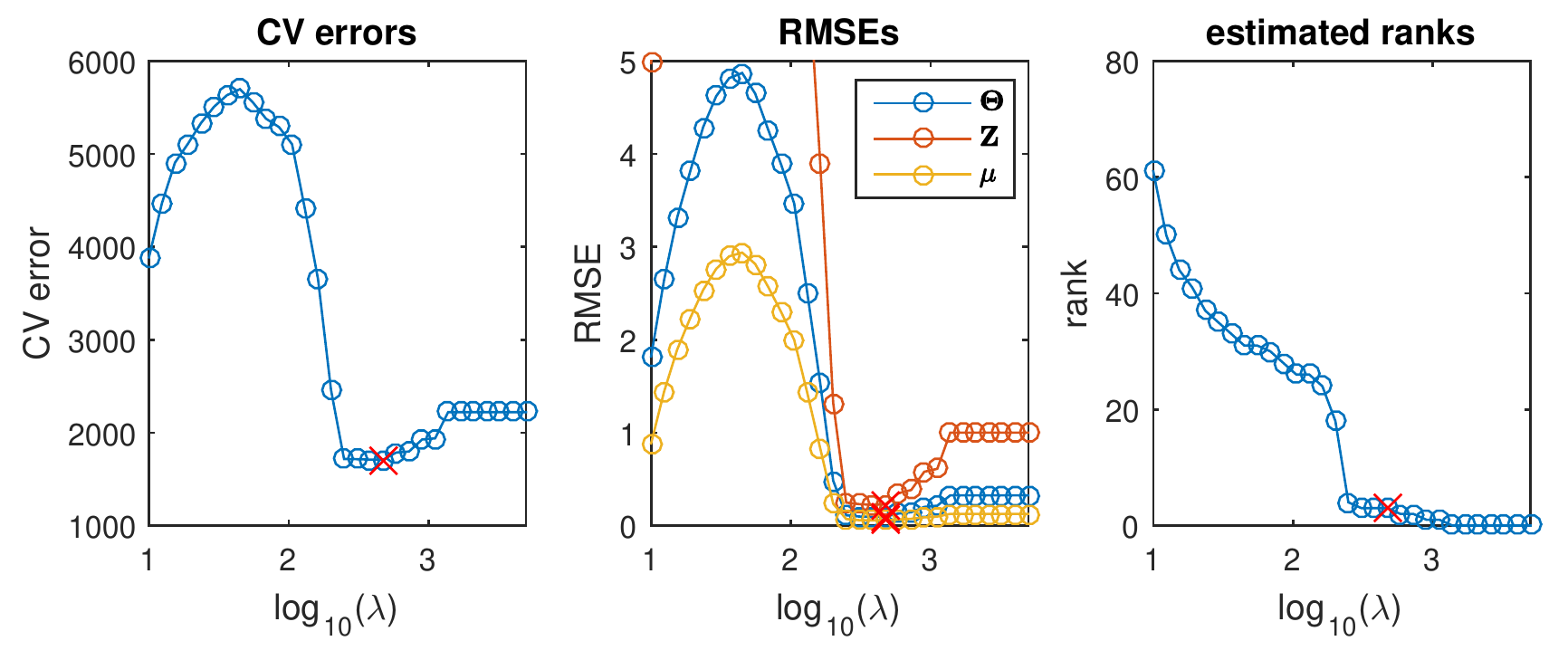}
    \caption*{\textbf{Fig.~3} Model selection and performance of the logistic PCA model with a GDP penalty. The CV error, RMSE of estimating $\mathbf{\Theta}$, $\mathbf{Z}$ and $\bm{\mu}$ and the estimated rank as a function of $\lambda$. The increased CV error and RMSEs for small $\lambda$ are the result of non-converged models after 500 iterations. The red cross marker indicates the $\lambda$ value where minimum CV error is achieved.}
\end{figure}

\begin{figure}[h!]\label{Fig:4}
    \centering
    \includegraphics[width=\textwidth ]{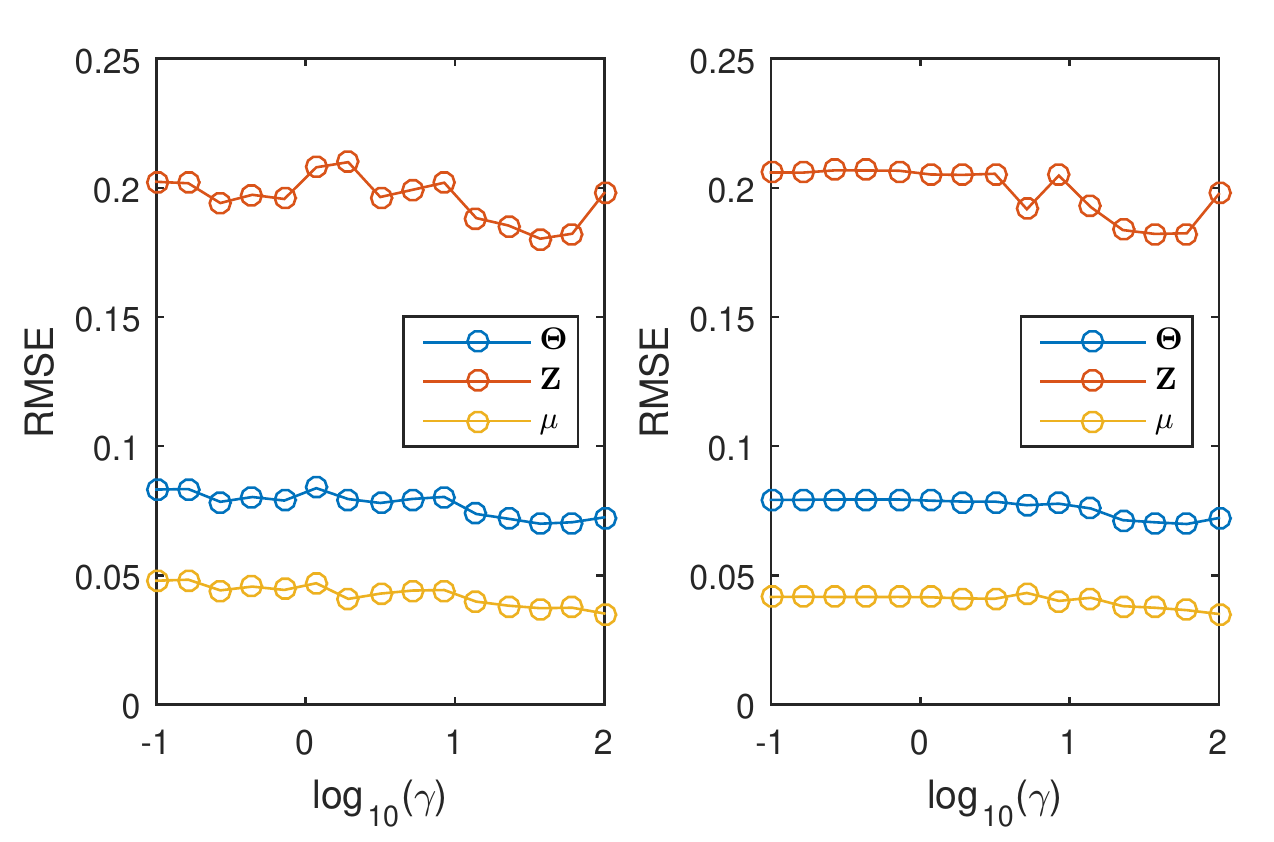}
    \caption*{\textbf{Fig.~4} The RMSE of estimating $\mathbf{\Theta}$, $\mathbf{Z}$ and $\bm{\mu}$ as a function of hyper-parameter $\gamma$. Results on the left side are obtained when the optimal model is selected based on minimum CV error while on the right hand side model selection was based on minimum  $\text{RMSE}(\mathbf{\Theta})$.}
\end{figure}

\subsection{The performance of the logistic PCA model using different penalties}
In this section, we compare the performance of the logistic PCA models with the exact low rank constraint, the nuclear norm penalty, and the GDP ($\gamma=1$) penalty. Random initialization is used and the maximum number of iterations is set to 10000 for all the models. Furthermore, all models are fitted using both $\epsilon_f = 10^{-6}$ and $\epsilon_f = 10^{-8}$ to test the model's robustness to the stopping criteria. For the standard logistic PCA model using the exact low rank constraint, 5 components are used. For the model with a GDP penalty, the above selected model is used, while for the model with nuclear norm penalty, the model is selected (the model selection results are shown in Fig.~S2) and re-fitted on full data set in the same way as was described above. In addition, according to the latent variable interpretation of the logistic PCA model, the unobserved quantitative data set $\mathbf{X}^{\ast} = \mathbf{\Theta} + \mathbf{E}$ is available in our simulation. We constructed a 5 components PCA model (with offset term) on this latent data $\mathbf{X}^{\ast}$, and this model is called the full information model. The results of above experiment are shown in Tab.~1. Since the logistic PCA model with nuclear norm penalty is a convex problem, the global solution can be achieved. The results from this model are taken as the baseline to compare other approaches. The drawback of the model with the nuclear norm penalty is that the proposed CV error based model selection procedure tends to select a too complex model to compensate for the biased estimation caused by the nuclear norm penalty (Fig.~S2, Tab.~1). The standard logistic PCA model with the exact low rank constraint can only achieve a relatively good estimation of the probability matrix $\mathbf{\Pi}$. The model is not robust to the stopping criteria. The logistic PCA model with a GDP penalty performs well in estimating the simulated parameters. Its results are even close to the full information model.\\

\begin{table}[htbp]
\centering
\caption*{Table 1: Comparison of the logistic PCA models with the exact low rank constraint, the nuclear norm penalty, the GDP penalty, and the full information model. The RMSEs of estimating $\mathbf{\Theta}$, $\mathbf{Z}$ and $\bm{\mu}$, as well as the mean Hellinger distance (MHD) of estimating the simulated probability matrix $\mathbf{\Pi}$ and the rank estimation of $\hat{\mathbf{Z}}$ are shown in the table.}
\label{Tbale:1}
\begin{tabular}{lllllll}
  \toprule
penalty & $\epsilon_f$ & $\text{RMSE}(\mathbf{\Theta})$ & $\text{RMSE}(\mathbf{Z})$ & $\text{RMSE}(\bm{\mu})$ & MHD & rank \\
  \midrule
  \multirow{2}{0.5em}{exact} & $10^{-6}$  &3.8017    &7.8491    &2.6023    &0.0726    &5      \\
                             & $10^{-8}$  &8.4955    &17.0129   &5.9715    &0.0733    &5      \\
  \hline
  \multirow{2}{0.5em}{nuclear norm} & $10^{-6}$  &0.1407    &0.3788    &0.0701    &0.0670   &27      \\
                                    & $10^{-8}$  &0.1405    &0.3783    &0.0700    &0.0670   &27     \\
  \hline
  \multirow{2}{0.5em}{GDP} & $10^{-6}$   &0.0797    &0.2064    &0.0421    &0.0515    &3    \\
                           & $10^{-8}$   &0.0786    &0.2063    &0.0408    &0.0514    &3    \\
  \hline
  full &   &0.0120    &0.0465    &0.0017    &0.0258    &5     \\

  \bottomrule
\end{tabular}
\end{table}

The difference in the performance of the logistic PCA models with different penalties (Tab.~1) are mainly related to how these penalties shrink the singular values. Therefore, we also compared the singular values of the simulated $\mathbf{Z}$ and their estimations from the logistic PCA models with different penalties, and its estimation from the full information model. The results are shown in Fig.~5. The simulated low rank is 5, however the last component is overwhelmed by the noise. Furthermore, the $4$-th component is less than 2 times noise level and therefore cannot be expected to be distinguished from the noise. From Fig.~5(left) it becomes clear that the standard logistic PCA model clearly overestimates the singular values of $\mathbf{Z}$. And when the more strict stopping criteria is used, the overestimation problem becomes even worse. Fig.~5(right) shows that the logistic PCA model with nuclear norm penalty underestimated the singular values of $\mathbf{Z}$, and includes too many small singular values into the model. The logistic PCA model with GDP penalty has very accurate estimation of the first three singular values of $\mathbf{Z}$. These results are in line with the their performance measures in Tab.~1 and their thresholding properties in Fig.~1.\\

\begin{figure}[h!]\label{Fig:5}
    \includegraphics[width = \textwidth]{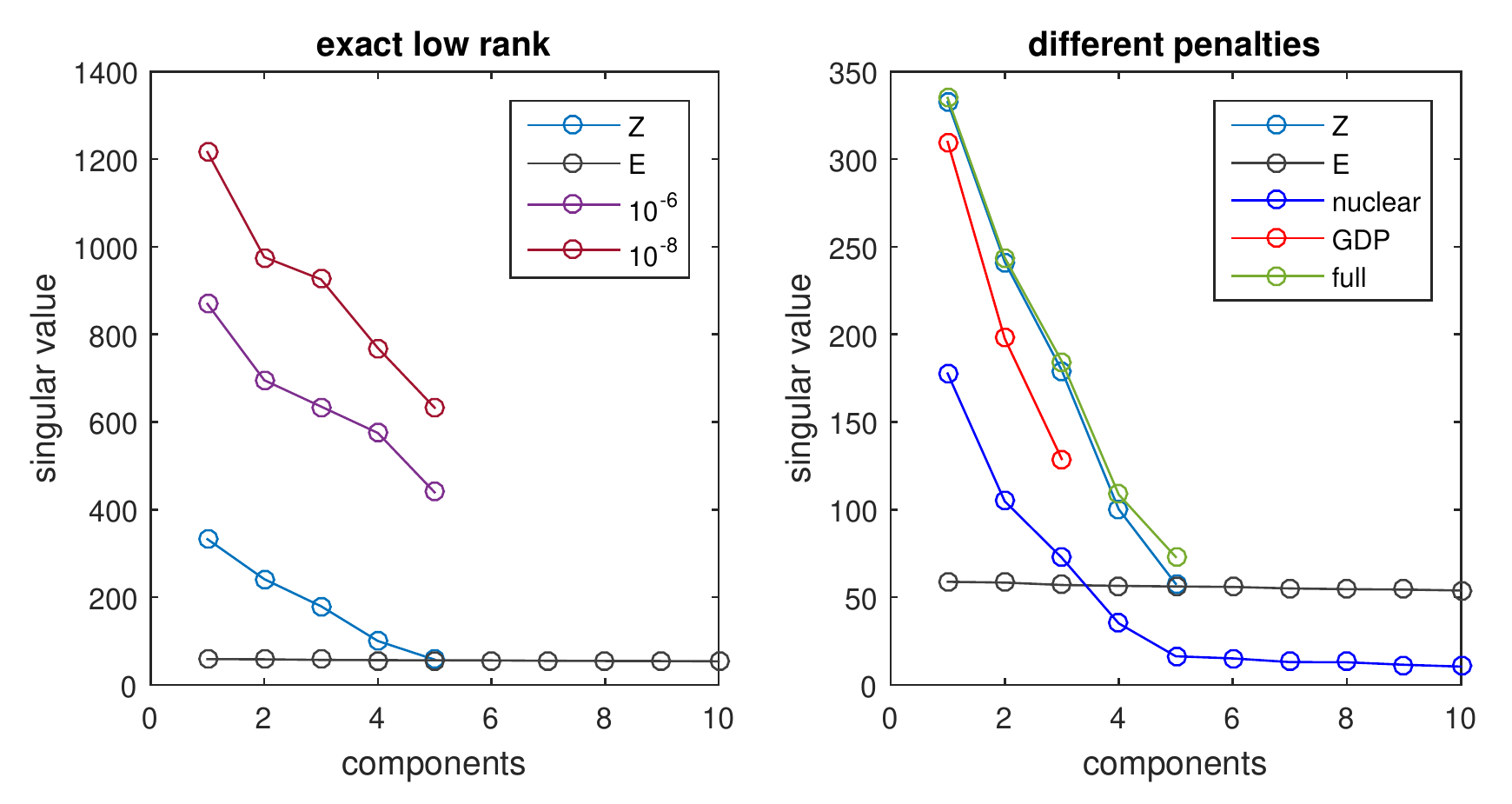}
    \caption*{\textbf{Fig.~5} Left: the singular values of the simulated $\mathbf{Z}$ and $\mathbf{E}$, and the singular values of the estimated $\hat{\mathbf{Z}}$ from a 5 components standard logistic PCA models ($\epsilon_f = 10^{-6}$ and $\epsilon_f = 10^{-8}$). Right: the singular values of the simulated $\mathbf{Z}$ and $\mathbf{E}$, and the singular values of the estimated $\hat{\mathbf{Z}}$ from the logistic PCA models ($\epsilon_f = 10^{-6}$) with a nuclear norm penalty and a GDP penalty, and from the full information model. Only the first 10 components are shown.}
\end{figure}

\subsection{Performance of the logistic PCA model as a function of SNR in the binary simulation.}
In the analysis of simulated quantitative data set using the PCA model, an increase in SNR makes the estimation of the true underlying low rank structure easier. Unfortunately, this is not true in the estimation of the true underlying logistic PCA model for simulated binary data. To illustrate this, the following experiment was performed. 30 SNR values are selected from the interval $[10^{-2}, 10^{3}]$ equidistant in log-space. The simulated offset term $\bm{\mu}$ is set to $\mathbf{0}$ to simulate balanced binary data sets, the number of samples, variables, and the low rank are the same as the experiment described above. For the binary data simulations with different SNRs, only the constant $c$, which is used to adjust the SNR, changes with the SNR. All other parameters are kept the same. For each simulated $\mathbf{X}$ with a specific SNR, logistic PCA models with GDP ($\gamma=1$) penalty and with nuclear norm penalty are selected and re-fitted. In addition, PCA models with different numbers of components are fitted on the latent quantitative data set $\mathbf{X}^{\ast}$, and the model with the minimum $\text{RMSE}(\mathbf{\Theta})$ is selected. In addition, the null model, i.e. the logistic PCA model with only the column offset term, is used to provide a baseline for comparison of the different approaches. The above experiments are repeated 10 times, and their results are shown in Fig.~6. Results obtained from a similar experiment but performed on imbalanced data simulation are shown in Fig.~S3. There the simulated $\bm{\mu}$ is set according to the marginal probabilities of the CNA data set. Overall, the logistic PCA models with different penalties can always achieve better performance than the null model, and the model with a GDP penalty demonstrate superior performance with respect to all the used metrics compared to the model with convex nuclear norm penalty.\\

Fig.~6 (left and center) shows that with increasing SNR, the estimation of the quantitative full model improves as expected. However, for the parameters estimated from the binary data this is not the case. First the estimation of the simulated parameters $\mathbf{\Theta}$ and $\mathbf{Z}$ improves, but when the SNR increases even further, the estimation deteriorates again leading to a bowl shaped pattern. This pattern has been observed before in a binary matrix completion using nuclear norm penalty \cite{davenport20141}. What is going on here? When the SNR is very low, elements in the simulated $\mathbf{\Theta}$ are close to 0, therefore elements in the simulated probability matrix $\mathbf{\Pi} = \phi(\mathbf{\Theta})$ are close to $0.5$, and the binary observations in $\mathbf{X}$ are simply random data without an underlying low rank structure. The logistic PCA model will not have good performance on such data sets.
When the SNR becomes very large, $\text{E}(x_{ij}|\theta_{ij}) = \phi(\theta_{ij})$, in which $x$ is $ij$-th term of $\mathbf{X}$ and $\theta$ is the $ij$-th term of $\mathbf{\Theta}$. The logistic PCA model aims to maximize the likelihood, $Pr(x_{ij}=1|\theta_{ij})=\phi(\theta_{ij})$. However, as can be seen in (Fig.~S4), the slope of $\phi(\theta_{ij})$ is almost flat when $\theta_{ij}$ is large. There is almost no difference in $\phi(\theta_{ij})$ if $\phi(100)$ or if $\phi(1000)$. Thus the model is able to reproduce the simulated $\mathbf{\Pi}$ based on the logistic PCA model almost exactly (Fig.~6 right), but the estimation of $\mathbf{\Theta}$ and $\mathbf{Z}$ are not accurate (Fig.~6 left and center). \\

\begin{figure}[h!]\label{Fig:6}
    \centering
    \includegraphics[width = \textwidth]{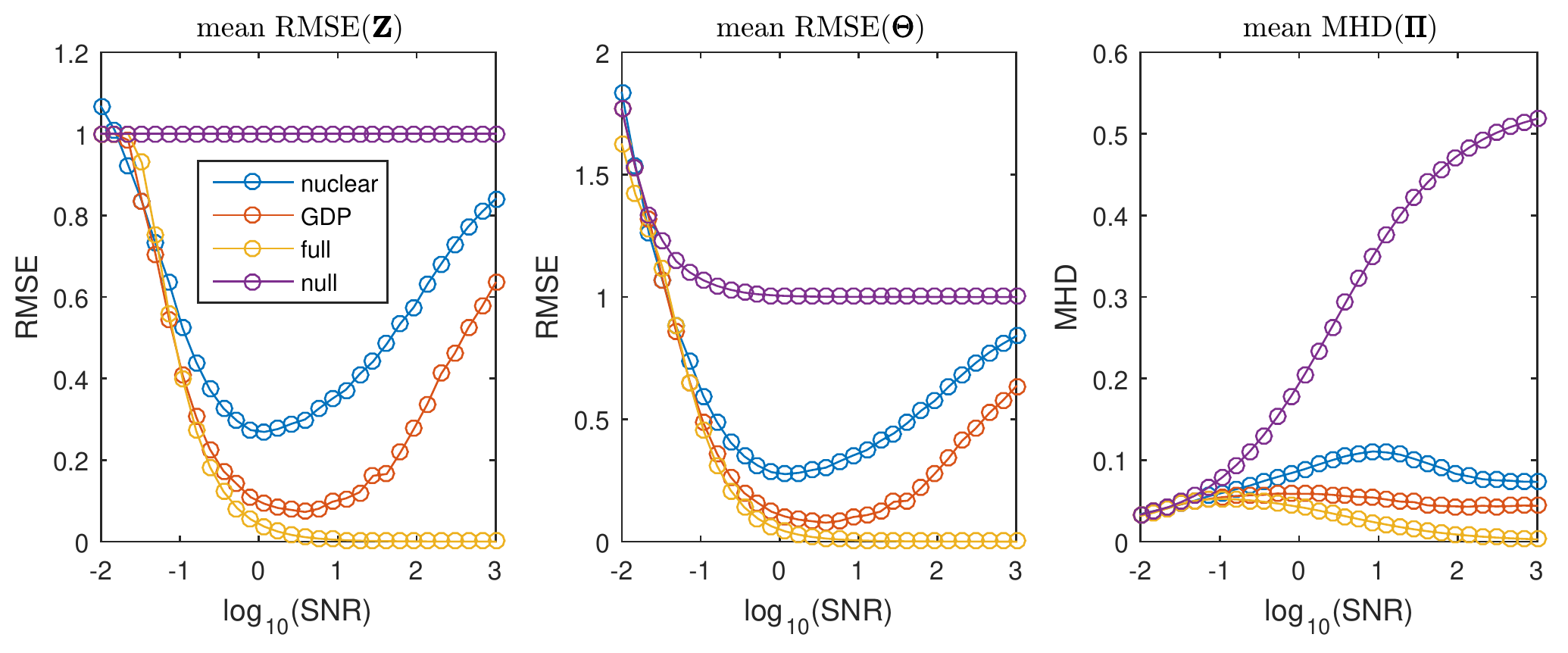}
    \caption*{\textbf{Fig.~6} RMSE of $\mathbf{Z}$ (left) and $\mathbf{\Theta}$ (middle) and the MHD of $\mathbf{\Pi}$ as a function of increasing SNR values for simulated balanced binary data.}
\end{figure}

\subsection{Real data analysis}
We demonstrate the proposed logistic PCA model with a GDP ($\gamma=1$) penalty and the corresponding model selection procedure on the CNA data set. The model selection is done in the same way as was described above, and the result is shown in Fig.~S5. After that, the selected 4 components model is re-fitted on the full data set. The score and loading plot of the first 2 components are shown in Fig.~7, and the variation explained ratios of the derived 4 components are shown in Fig.~S6. As was explained before in \cite{song2018generalized}, the CNA data set is not discriminative for the three cancer types (illustrated in the score plot of Fig.~7 left). The structure in the loading plot (Fig.~7 right) mainly explains the technical characteristics of the data. Fig.~7 (right) shows that the gains and losses of the segments in the chromosomal regions corresponding to the CNA measurements are almost perfectly separated from each other in the first component. Therefore, the 63.05\% variation explained of the first component (Fig.~s6) is mainly because of the difference of gains and losses in CNA measurements. The remainder of the explained variation is rather limited.\\

\begin{figure}[h!]\label{Fig:7}
    \centering
    \includegraphics[width = \textwidth]{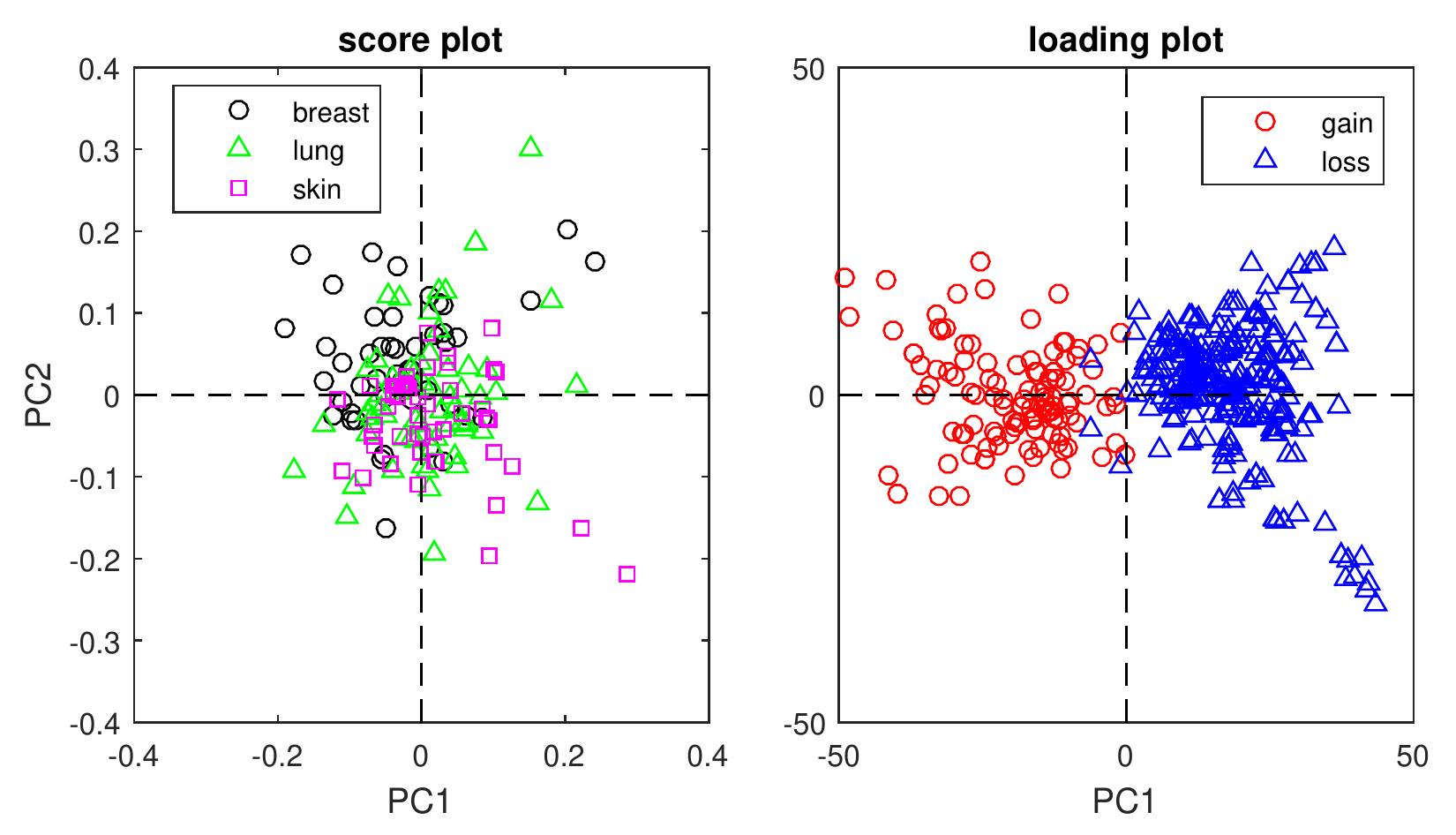}
    \caption*{\textbf{Fig.~7} The score and loading plots of the first 2 components of the logistic PCA model on the CNA data. The legend, breast, lung and skin, indicates the corresponding three cancer types. The legend, gain, loss, indicates the gain or loss of a segment in the chromosome region corresponding to the CNA measurement.
    }
\end{figure}

\section{Discussion}
To study the properties of the logistic PCA model with different penalties, we need to have the ability to simulate the multivariate binary data set with an underlying low rank structure, and the simulated structure should have a proper SNR so that the model can find it back. The latent variable interpretation of the logistic PCA model not only makes the assumption of low rank structure easier to understand, but also provides us a way to define SNR in multivariate binary data simulation.\\

The standard logistic PCA model using the exact low rank constraint has an overfitting problem. The overfitting issue manifests itself in a way that some of the elements in the estimated loading matrix $\hat{\mathbf{B}}$ (the orthogonality constraint is imposed on $\mathbf{A}$) have the tendency to approach to infinity, and the non-zero singular values of the $\hat{\mathbf{Z}} = \hat{\mathbf{A}}\hat{\mathbf{B}}^{\text{T}}$ are not upper-bounded when strict stopping criteria are used. This overfitting issue can be alleviated by regularizing the singular values of $\mathbf{Z}$. Both convex nuclear norm penalties and concave GDP penalties can induce low rank estimation and simultaneously constrain the scale of the non-zero singular values. Therefore, logistic PCA models with these penalties do not suffer from the overfitting problem.\\

However, the logistic PCA model with a GDP penalty has several advantages compared to the model with the nuclear norm penalty. Since the nuclear norm penalty applies the same degree of shrinkage on all the singular values, the large singular values are shrunken too much. Therefore, the implemented CV error based model selection procedure tends to select a very complex model with too many components to compensate for the biased estimations. On the contrary, GDP penalty achieves nearly unbiased estimation. Thus the CV error based model selection is successful in selecting the logistic PCA model with the a GDP penalty. Furthermore, the selected logistic PCA model with GDP penalty has shown superior performance in recovering the simulated low rank structure compared to the model with the nuclear norm penalty, and the exact low rank constraint.\\

\bibliographystyle{ieeetr}

\bibliography{reference}

\newpage
\section*{Supplementary files}
\section*{Supplementary figures}
\begin{figure}[h!]\label{Fig:S1}
    \centering
    \includegraphics[width=\textwidth ]{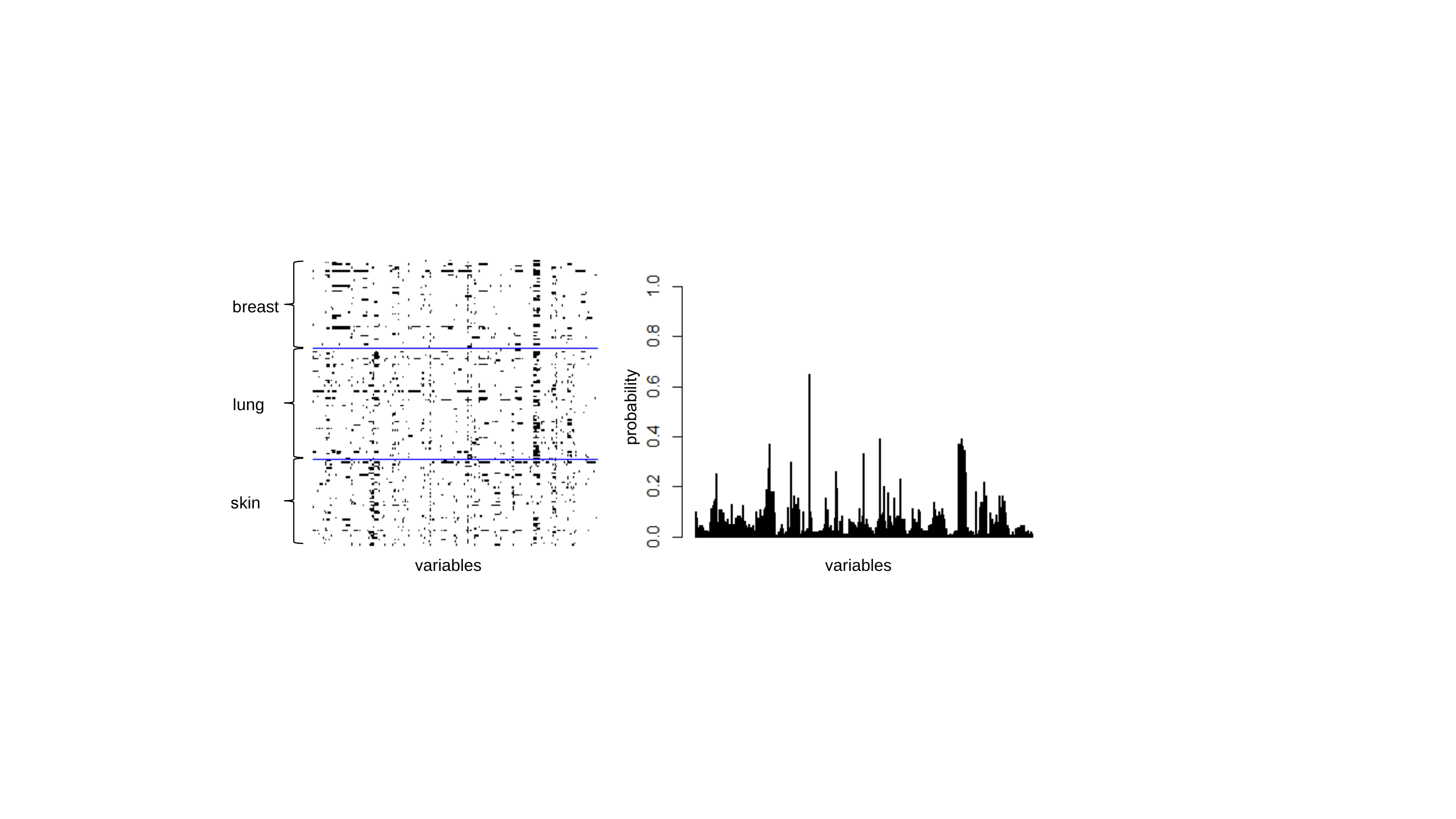}
    \caption*{\textbf{Fig.~S1} Left: the heat map of the CNA data, in which black color indicates ``1'' and white color, ``0''. Right: the empirical marginal probabilities of the CNA variables.}
\end{figure}

 \begin{figure}[h!]\label{Fig:S2}
    \centering
    \includegraphics[width= \textwidth ]{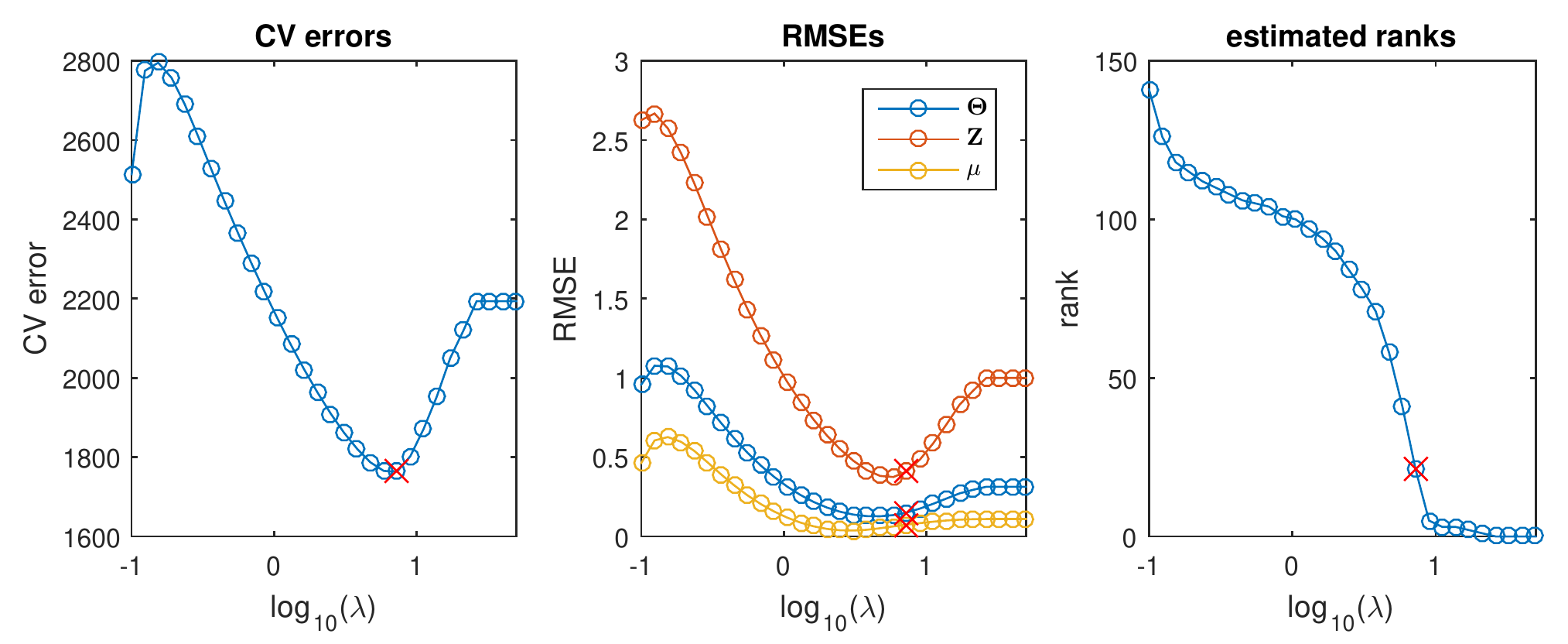}
    \caption*{\textbf{Fig.~S2} Model selection and performance of the logistic PCA model with the nuclear norm penalty. The CV error, RMSE of estimating $\mathbf{\Theta}$, $\mathbf{Z}$ and $\bm{\mu}$ and the estimated rank as a function of $\lambda$. The increased CV error and RMSEs for small $\lambda$ are the result of non-converged models after 500 iterations. The red cross marker indicates the $\lambda$ value where minimum CV error is achieved.}
\end{figure}

\begin{figure}[h!]\label{Fig:S3}
    \centering
    \includegraphics[width = \textwidth]{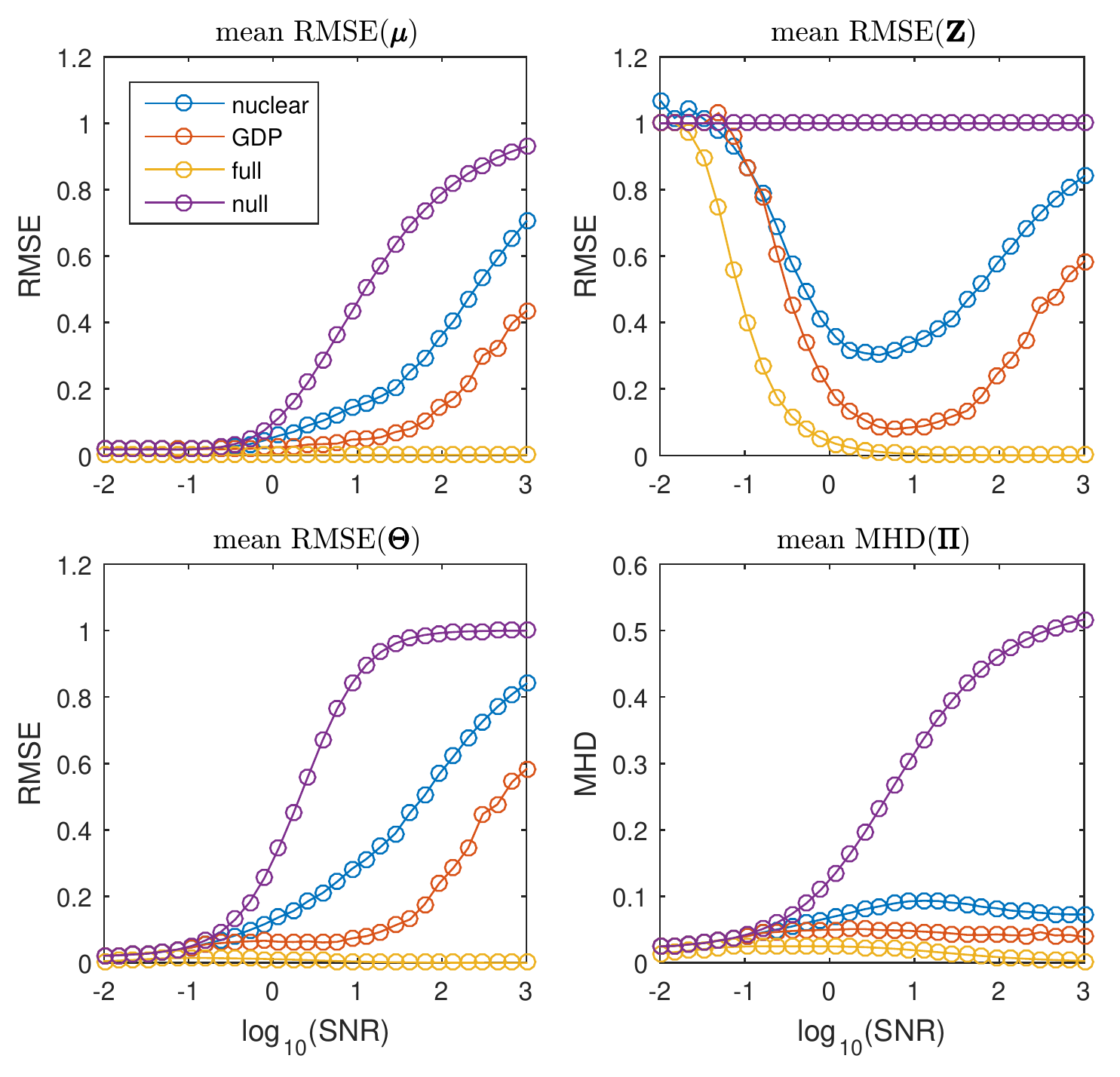}
    \caption*{\textbf{Fig.~S3} How the SNR in imbalanced binary data simulation affects the performance of the logistic PCA models with different penalties, and the full information model.}
\end{figure}

\begin{figure}[h!]\label{Fig:S4}
    \centering
    \includegraphics[width = 0.5\textwidth]{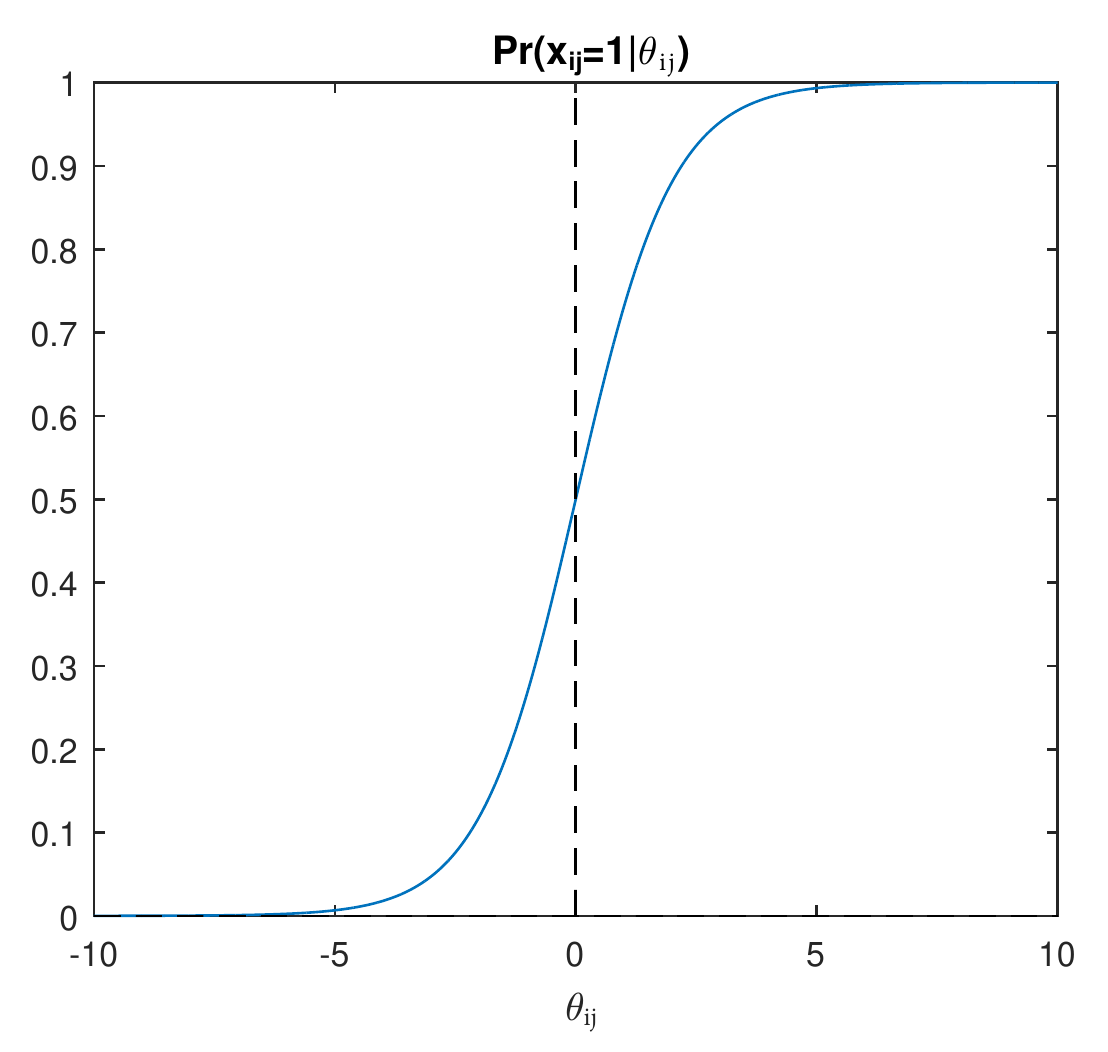}
    \caption*{\textbf{Fig.~S4} The relationship of $\text{E}(x_{ij}|\theta_{ij}) = \phi(\theta_{ij})$, in which $x_{ij}$ and $\theta_{ij}$ are the $ij$-th elements of $\mathbf{X}$ and $\mathbf{\Theta}$. }
\end{figure}

\begin{figure}[h!]\label{Fig:S5}
    \centering
    \includegraphics[width = \textwidth]{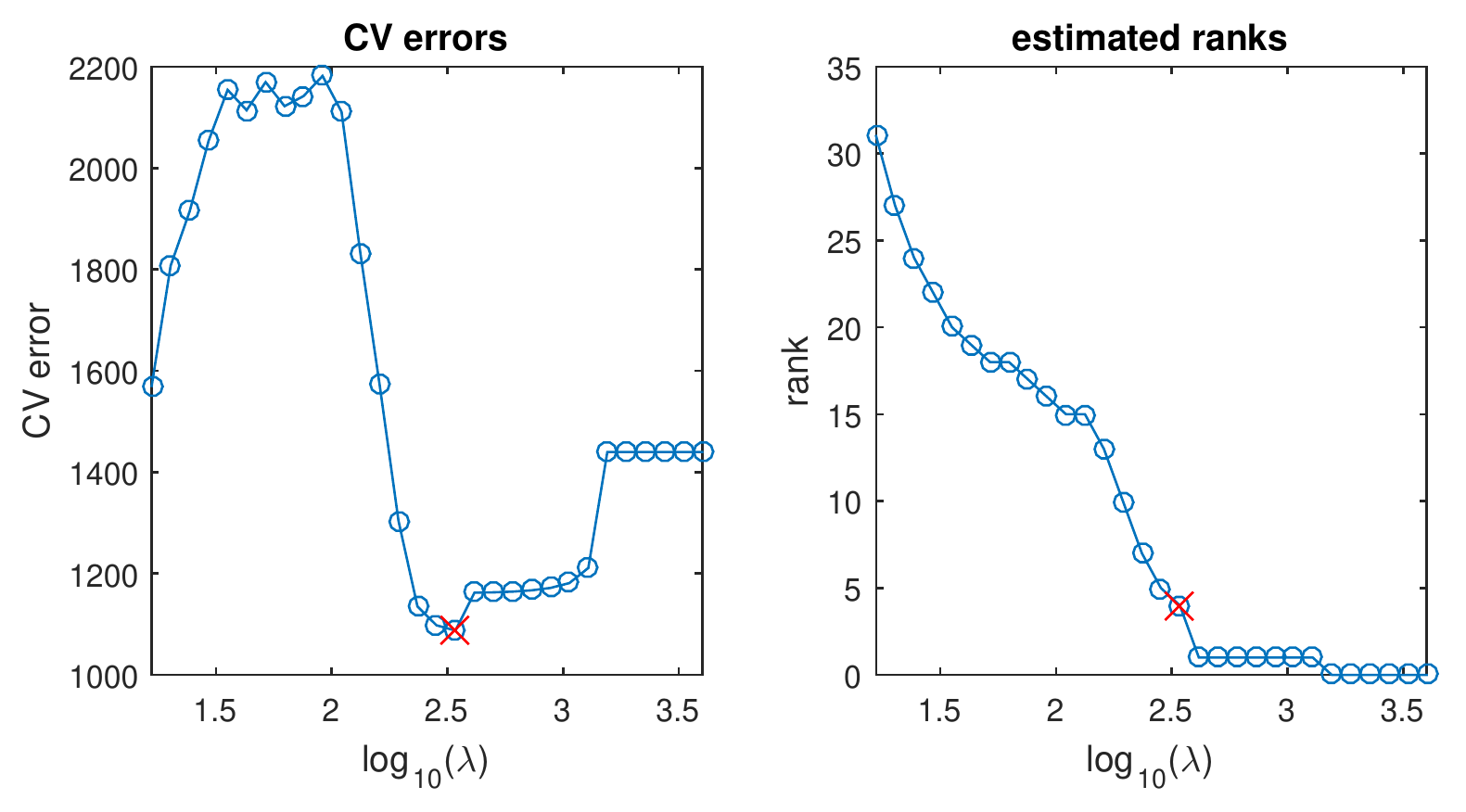}
    \caption*{\textbf{Fig.~S5} How $\lambda$ effects the CV error (left) and the rank estimation (right) in the model selection process of the logistic PCA model with a GDP penalty on the CNA data set.}
\end{figure}

\begin{figure}[h!]\label{Fig:S6}
    \centering
    \includegraphics[width = 0.5\textwidth]{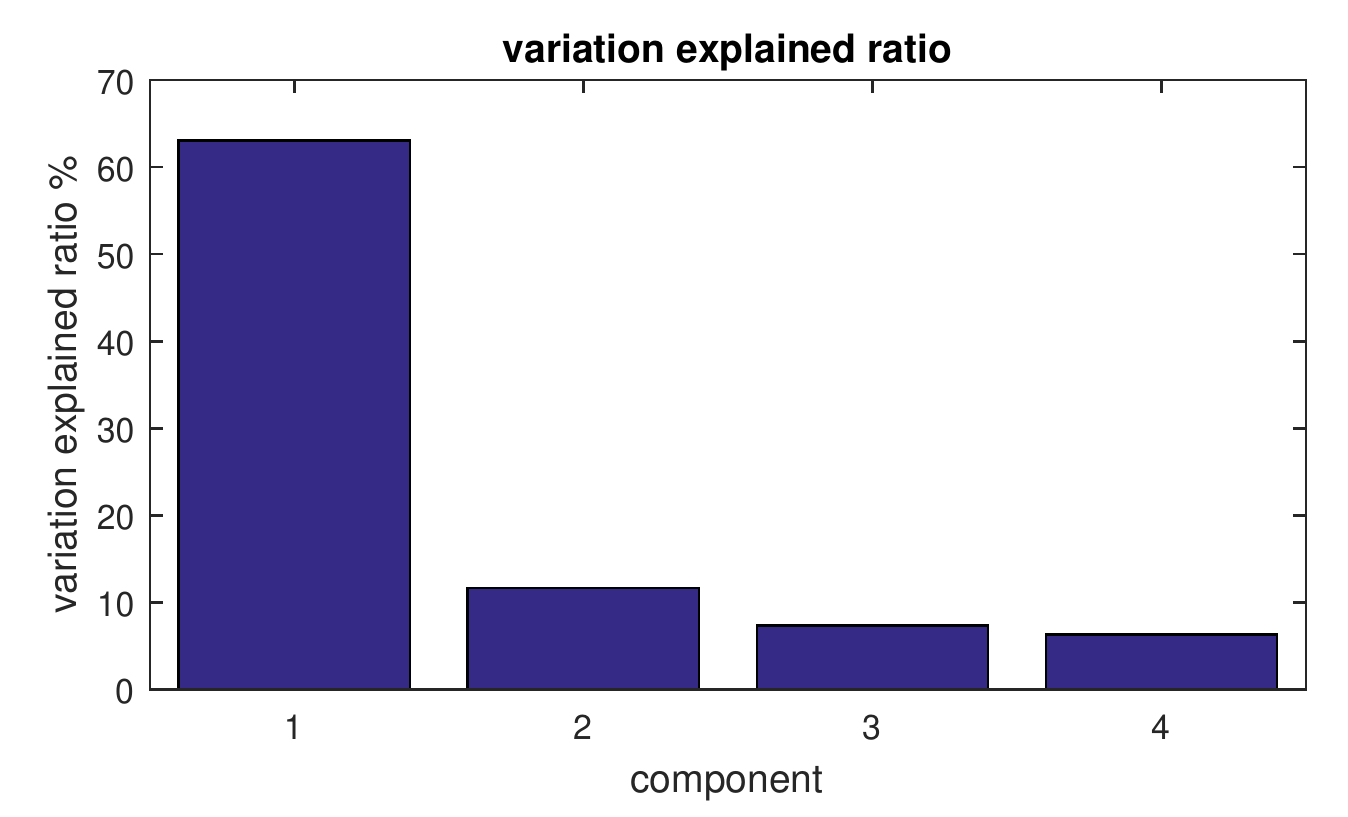}
    \caption*{\textbf{Fig.~S6} The variation explained ratios of the 4 components in the logistic PCA model on the CNA data set.}
\end{figure}

\end{document}